\documentclass{elsart1p}
\usepackage{graphicx}
\usepackage{amssymb}
\usepackage{color}
\def\gsim{\mathrel{\raise.3ex\hbox{$>$\kern-.75em\lower1ex\hbox{$\sim$}}}}

\def\lsim{\mathrel{\raise.3ex\hbox{$<$\kern-.75em\lower1ex\hbox{$\sim$}}}}
\begin{document}
\begin{frontmatter}
%
%
%
\title{Laboratory tests for the cosmic neutrino background using
beta-decaying nuclei}
%
%
\author{Bob McElrath}
\address{CERN, Geneva 23, CH-1211 Switzerland}
\begin{abstract}
    We point out that the Pauli blocking of neutrinos by
    cosmological relic neutrinos can be a significant effect.  For
    zero-energy neutrinos, the standard parameters for the neutrino
    background temperature and density give a suppression of
    approximately $1/2$.  We show the effect this has on three-body beta
    decays.  The size of the effect is of the same order as the recently
    suggested neutrino capture on beta-decaying nuclei.
\end{abstract}
\end{frontmatter}
%

\section{Introduction}

The last remaining remnant of the big bang, which is composed of a {\it
known} particle is the Cosmic Neutrino Background (CNB).  It decouples
from thermal equilibrium at $T \sim 2$ MeV.  It gives information about
the universe at a time significantly before the decoupling of photons at
$T \sim 1$ eV.  The processes responsible for their creation and
decoupling are well understood nuclear physics.

Today these neutrinos are expected to be extremely cold ($1.952 {\rm K}
= 1.68\times 10^{-4} {\rm eV}$).  As such, they are extremely difficult
to detect due to the fact that weak interaction cross sections scale as
$(G_F E)^2$.  They have a density of $\rho_\nu=3/22 n_\gamma = 56/{\rm
cm}^3$ per species of neutrino and anti-neutrino, corresponding to a
luminosity $\mathcal{L} = 1.7\times 10^{13}/{\rm cm^2s}$ if the
neutrinos were massless.~\cite{Ringwald:2005zf,Gelmini:2004hg}
These numbers rely on a specific cosmological model which could be
substantially modified, if neutrinos cluster
gravitationally, or if they have nontrivial dynamics after
freeze-out.~\cite{Ringwald:2004np}  It has also recently been shown that
these neutrinos are a quantum liquid, and their fluctuations have the
quantum numbers of a graviton, opening the prospect that measurements of 
relic neutrinos could then be compared with gravitational 
constants.~\cite{McElrath:2008ye}

Nuclei that undergo $\beta$-decay are a precise and specific laboratory to look
for the CNB.  There exists a vast array of nuclei that can emit or absorb
neutrinos at a wide range of energies.  A signal seen using $\beta$-decaying
nuclei constitutes a specific test because they are already known to emit or
absorb lepton number in specific ways.  All other proposals could not in
principle tell if the effect was due to an object with lepton
number.~\cite{Ringwald:2005zf,Gelmini:2004hg} High-energy neutrinos (e.g.
Z-burst) can be absorbed by things which do not carry lepton number, and
anomalous forces could have a variety of sources that have nothing to do with
lepton number (for instance, a Dark Matter wind).  Finally the effects of
neutrinos on the CMB cannot be disentangled from other relativistic species
that are not be fermionic or do not carry lepton number.

Rates using decaying nuclei are in principle much higher than other
proposals such as coherent scattering, because the energy for the
observation is coming from nuclear mass differences, and not from the
neutrinos themselves.  The energy $Q$ in nuclear transitions is
$\mathcal{O}({\rm MeV})$.  Giving the CNB neutrinos this energy in a
coherent experiment would require moving with a velocity corresponding
to a boost factor $\gamma = Q/T_\nu \simeq 10^{10}$.  For comparison,
$\gamma$ at the LHC with protons is about 15000, or 5500 with Lead.

There are two ways to see an effect of the neutrino background using a
beta-decaying nucleus: add a neutrino to it or remove a neutrino from it.  Both
were suggested by Weinberg in 1962.~\cite{Weinberg:1962zz} 

Adding a neutrino to the background is suppressed for momenta which are
already occupied by the CNB thermal distribution, due to the fact that
neutrinos are fermions and their chemical potential and average energy
are similar.  This is an $\mathcal{O}(1)$ effect, if one can create
neutrinos having the correct energy.

Removing a neutrino from the CNB using nuclei is known as neutrino
capture ($\nu$C).  Capture of reactor neutrinos is the original mode
used to discover the neutrino.  Recently there has been a surge of
interest in this mode for detecting the CNB using $\beta$-decaying
nuclei, which can have zero
threshold.\cite{Cocco:2007za,Lazauskas:2007da,Blennow:2008fh}


\section{Pauli Blocking by The Cosmic Neutrino Background}

The CNB is a thermal distribution in a particular frame $u^\alpha$ which
we assume to be coincident with the dipole from the Cosmic Microwave
Background, which points in the direction $(264.85 \pm 0.10)^\circ,
(48.25 \pm 0.04)^\circ$ in galactic coordinates, with velocity $368 \pm
2$ km/s.  Its thermal distribution is then
\begin{eqnarray}
    F_i(\vec{p}) = \left[e^{(p^\alpha u_\alpha - \mu_i)/kT} + 1 \right]^{-1}
    \label{eq:F}
\end{eqnarray}
for each species of neutrino and anti-neutrino $i$, having mass $m_i$
and chemical potential $\mu_i$ and four-momentum $p^\alpha$ in the
cosmic rest frame $u^\alpha=(1,\vec{0})$ and this reduces to the usual
non-relativistic Fermi-Dirac distribution.  The relativistic chemical
potential is the Fermi energy at zero temperature, $\mu_i = E_F =
\sqrt{m^2 + p_F^2}$, and the nominal Fermi momentum predicted
by the standard cosmological model is $p_F = \sqrt[3]{3 \pi^2 \rho_\nu}
= \sqrt[3]{3 \pi^2 \rho_{\overline{\nu}}} =3.6\times 10^{-5}$ eV, where
$\rho_\nu$ is the number density per flavor.  We will refer to this as
the ``standard'' chemical potential.  

\begin{figure}[ht]
    \begin{minipage}[b]{0.47\linewidth}
        \centering
        \includegraphics[scale=0.25]{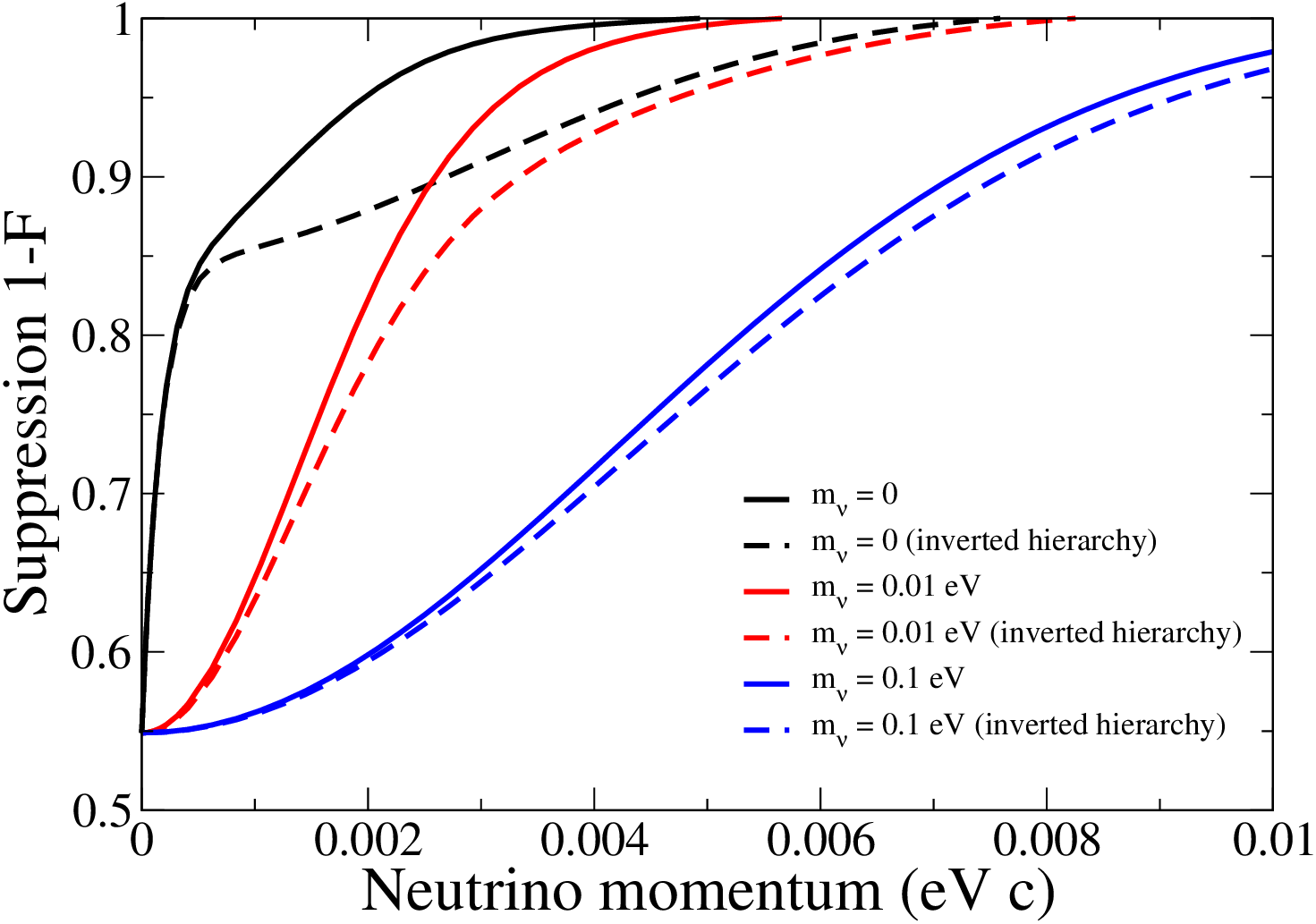}
        \caption{Sum of suppression factor $1-F_i(p)$ for three mass
        eigenstates vs. neutrino momentum for several values of the neutrino
        mass, assuming a standard chemical potential.}
        \label{fig:FDsmasses}
    \hspace{0.05cm}
    \end{minipage}
    \hspace{0.5cm}
    \begin{minipage}[b]{0.47\linewidth}
    \hspace{0.05cm}
        \centering
        \includegraphics[scale=0.25]{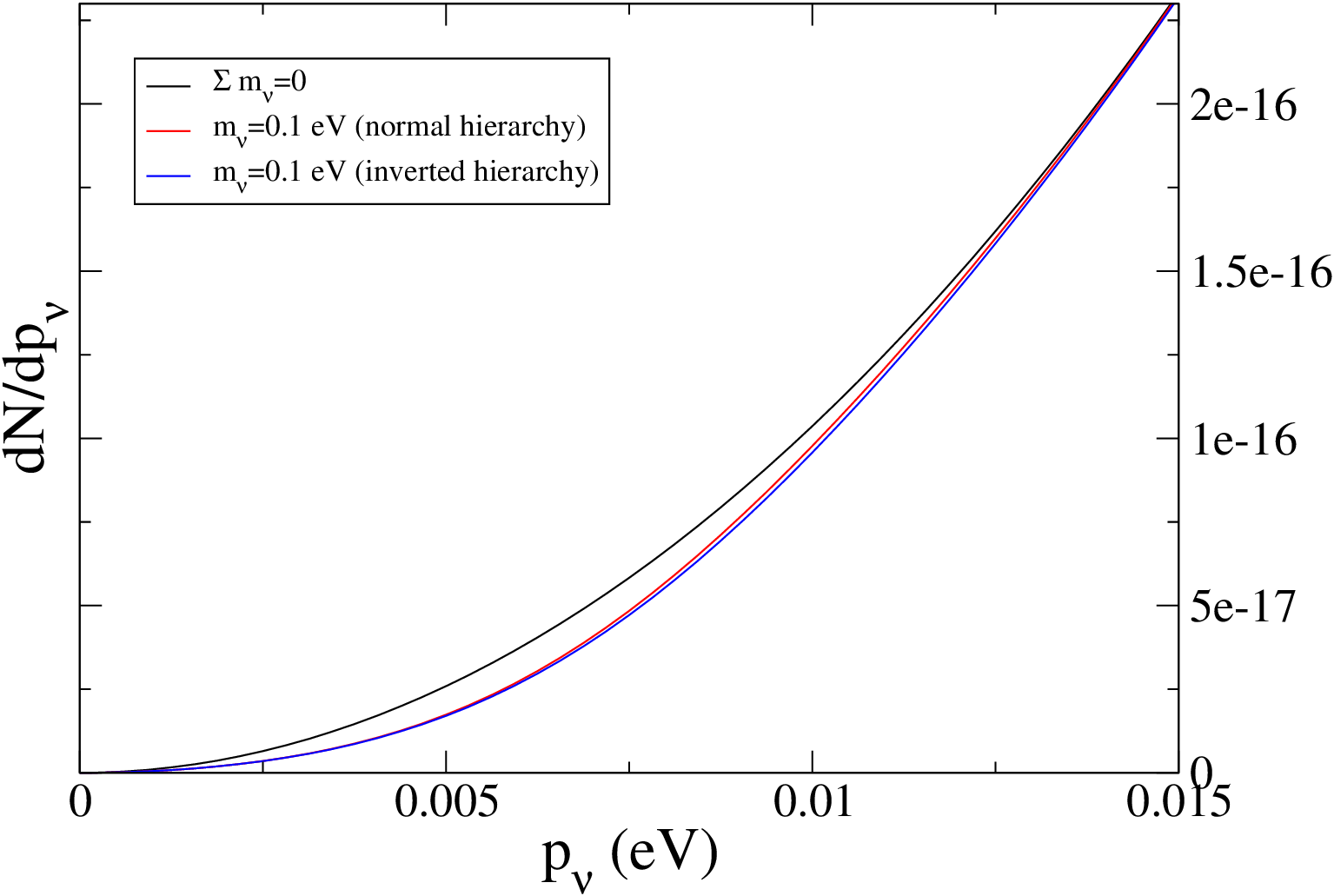}
        \caption{The differential event rate as a function of neutrino
        momentum for several choices of neutrino mass and normal/inverted
        hierarchy, assuming a standard chemical potential and the kinematics
        of Tritium.}
        \label{fig:decrate}
    \end{minipage}
\end{figure}

A process which emits neutrinos has a suppression $[1-F_i(\vec{p})]$ due
to Pauli blocking from this thermal distribution.  This is independent
of whether the neutrinos are described as a localized classical gas
having small uncertainty $\Delta x \ll n^{-1/3}$ or a quantum liquid
$\Delta x \gg n^{-1/3}$ for number density $n$.  For a beta decay this
is \begin{equation}
d \Gamma = 2 \pi \sum_i \int |\mathcal{M}_i|^2 \xi_i^2 [1-F_i(\vec{p})] dPS
\end{equation}
where $dPS$ is the differential phase space, $\xi_i$ is the eigenvector
component of neutrino mass eigenstate $i$ in the electron neutrino
direction, $\mathcal{M}$ is the matrix element of the emission of an
electron-type neutrino with mass $m_i$, and one sums over the mass
eigenstates since final state emitted particles must be in a mass
eigenstate.  This suppression factor is experimentally indistinguishable
from $1$ except in a region in which the emitted neutrino has the same
energy as the CNB.  We plot the suppression factor, summed over flavors
in Fig.\ref{fig:FDsmasses}


The Matrix Element for the beta decay is 
\begin{eqnarray}
    \nonumber
|\mathcal{M}|^2 = 
\frac{G_F^2}{128 \pi^3 M_I^2} \big[&&
(g_V+g_A)^2(p_J\cdot p_e)(p_I\cdot p_\nu) \\
\nobreak
    &+& (g_V-g_A)^2(p_I\cdot p_e)(p_J\cdot p_\nu)  \\
    &+& (g_V^2-g_A^2)(p_I\cdot p_J)(p_e\cdot p_\nu)\big]
    \nonumber
\end{eqnarray}
where $g_V$ and $g_A$ are the vector and axial vector weak charges of
the atom.  For $I$=neutron, $g_V = g_A = -1/2$.  The matrix element also
reaches a minimum $|\mathcal{M}|^2=0$ at $p_\nu$=0, so the event rate at
$p_\nu=0$ is zero.  Because of this, the minimum in
Fig.\ref{fig:FDsmasses} is deceptive.  The differential event rate
including the matrix element is plotted in Fig.\ref{fig:decrate} for
tritium, assuming a standard chemical potential.

To place a neutrino into the background with significant suppression, we
need that the invariant $p^\alpha u_\alpha/kT_\nu \lsim E_F/kT_\nu$.
Since our velocity with respect to the cosmic rest frame is small
($\beta = \frac{v}{c} = 1.23\times 10^{-3}$), one can ignore our
velocity and pretend we can do the experiment in the cosmic rest frame.
\footnote{One must use Special relativity and not Galilean relativity
here.  The atoms and electron may be non-relativistic, but the neutrino
is relativistic.}

In a normal $\beta^\pm$ decay an atom $I$ decays to atom $J$ by emitting
an electron or positron and an anti-neutrino or neutrino.  If one can
precisely measure the momenta of $I$, $J$, and $e^\pm$, one can solve
for the neutrino momenta.  This requires momentum resolution on each of
order $\delta p \lsim \sqrt{2 m k T_\nu}$.  If the initial state $I$ is
at rest, this corresponds to a temperature 
\[ 
    T = \frac{2}{3}\frac{m_\nu}{M_I} T_\nu \simeq
    1.40\times 10^{-9} \left(\frac{m_\nu}{\rm eV}\right)\left(\frac{\rm
    amu}{M_I}\right) \rm K.
\]
Stated another way, the de Broglie wavelength of the neutrinos is 1.2
mm.  Since the uncertainty on momentum scales with momentum, the initial
state must have a similar de Broglie wavelength.  Modern atomic
Bose-Einstein Condensate and Degenerate Fermi Gas experiments using
laser and evaporative cooling routinely reach $10^{-9}$ K today.
Another promising technology to get to these precisions in the initial
state is ``Crystallized Beams''.~\cite{dan02}

The final state of the decay is the atom $J$ almost exactly back-to-back
with the electron or positron.  Similarly these final state particles
must be measured with a precision 
\[
\frac{\delta p}{p} \simeq \sqrt{\frac{2 m k T_\nu}{Q(Q+2 m_e)}}
\simeq 1.33\times 10^{-7} \sqrt{\frac{m_\nu}{eV}} \sqrt{\frac{18 {\rm
keV}}{Q}}.
\]
where in the last term we assume $Q < 2 m_e$.

One might wonder if the effect shown here can impact experiments such as
KATRIN which attempt to measure the neutrino mass using the highest
energy electrons in a beta decay.  KATRIN attempts to measure mass due
to the change in slope and rate suppression near the endpoint, and they
do not have the resolution to see the actual endpoint itself.  Their
resolution is approximately $\Delta E_e \sim 1$ eV.  The effects here
only effect the highest (electron) energy bin, and reduce the number of
events there by $\mathcal{O}(10^{-18}N)$ where $N$ is the total decays
they see.  Existing experiments simply do not have the rate for this
effect to be a concern.

\section{Acknowledgements}
We thank Patrick Huber and Mats Landroos for fruitful discussions.

%
%
%

%

\begin{thebibliography}{00}

\bibitem{Ringwald:2005zf}
  A.~Ringwald,
  arXiv:hep-ph/0505024.

\bibitem{Gelmini:2004hg}
  G.~B.~Gelmini,
  Phys.\ Scripta {\bf T121} (2005) 131
  [arXiv:hep-ph/0412305].
\bibitem{Ringwald:2004np}
  A.~Ringwald and Y.~Y.~Y.~Wong,
  JCAP {\bf 0412} (2004) 005
  [arXiv:hep-ph/0408241].

\bibitem{McElrath:2008ye}
  B.~McElrath,
  arXiv:0812.2696 [gr-qc].

%
\bibitem{Weinberg:1962zz}
  S.~Weinberg,
  Phys.\ Rev.\  {\bf 128} (1962) 1457.

\bibitem{Cocco:2007za}
  A.~G.~Cocco, G.~Mangano and M.~Messina,
  JCAP {\bf 0706} (2007) 015
  [arXiv:hep-ph/0703075].

\bibitem{Lazauskas:2007da}
  R.~Lazauskas, P.~Vogel and C.~Volpe,
  J.\ Phys.\ G {\bf 35} (2008) 025001
  [arXiv:0710.5312 [astro-ph]].

\bibitem{Blennow:2008fh}
  M.~Blennow,
  arXiv:0803.3762 [astro-ph].

\bibitem{dan02} H. Danared, A. K\"allberg, K.-G. Rensfelt, and A.
    Simonsson, Phys. Rev. Lett. 88, 174801 (2002).


%
%
%
%
\end{thebibliography}
\end{document}